\documentclass[aps,prl,twocolumn,superscriptaddress,showpacs]{revtex4-1}
\usepackage{graphicx}
\usepackage{bm}
\usepackage{amsmath, amsthm, amssymb}
\usepackage{color}

\newcommand{\ve}[1]{\mathbf{#1}}
\newcommand{\vk}{\ve{k}} % Vector k

\newcommand{\su}{\uparrow}    % Make the code more readable...
\newcommand{\sd}{\downarrow}  % Make the code more readable...
\newcommand{\e}[1]{\mathrm{e}^{#1}}
\newcommand{\etal}{\emph{et al.}}
\def\i{\mathrm{i}}

\newcommand{\kmpar}{\lambda_{\text{SO}}}
\newcommand{\stpar}{\lambda_{\text{v}}}

\renewcommand{\section}[1]{{\par\it #1.---}\ignorespaces}

\usepackage{soul}

\begin{document}

\title{Quantum ground state control in superconductor-silicene structures: 0-$\pi$ transitions, $\varphi_0$-junctions, and Majorana bound states}

\author{Dushko Kuzmanovski}
\affiliation{Department of Physics and Astronomy, Uppsala University, Box 516, SE-751 20 Uppsala, Sweden}

\author{Jacob Linder}
\affiliation{Department of Physics, NTNU, Norwegian University of
Science and Technology, N-7491 Trondheim, Norway}

\author{Annica Black-Schaffer}
\affiliation{Department of Physics and Astronomy, Uppsala University, Box 516, SE-751 20 Uppsala, Sweden}

\begin{abstract}
We demonstrate theoretically that proximity-induced superconductivity in silicene offers the possibility to exert strong quantum ground state control. We show that electrically controlled $0$-$\pi$ transitions occur in Josephson junctions in the presence of an exchange field due to the buckling of the silicene lattice. We also discover that zigzag-oriented interfaces, featuring intervalley scattering, cause a $\varphi_0$ state with an applied electric field. Finally, we demonstrate that Majorana bound states along the silicene edge are tunable via the edge orientation, electric, and in-plane spin exchange fields.
\end{abstract}

\date{\today}


\maketitle

% ===================== Introduction - BEGIN ================================
% Silicene intro:
The discovery of new low-dimensional materials, where the electron bands have topological properties, has attracted a large amount of interest in recent years. A particularly intriguing material is silicene~\cite{silicene_review}, which consists of an atomically thin, buckled layer of Si atoms arranged in a honeycomb lattice. This material has theoretically been shown to host both different topological phases and has individually tunable mass gaps for each spin $\sigma$ at each valley $\eta$~\cite{liu_prl_11,ezawa_njp_12,drummond_prb_12,ezawa_prl_12}. These properties make silicene ideal for envisaging various types of device functionality related to both spintronics and valleytronics, a quest also significantly fueled by its compatibility with existing Si semiconductor technology. On the experimental side, silicene has already been studied on metallic substrates, including ZrB$_2$~\cite{fleurence_prl_12}  and notably Ag(111)~\cite{chiappe_advmater_12,feng_nanolett_12,vogt_prl_12}, as well as for nonmetallic hosts, where a Si nanosheet grown on MoS$_2$ bulk crystals has recently been reported~\cite{chiappe_advmater_14}.

% Superconducting silicene
A particularly exciting prospect is to consider the manifestation of superconducting correlations in silicene, with the unique properties of silicene likely leading to an advanced interplay between spintronics and superconductivity~\cite{linder_nphys_15}. Recent experimental progress has enabled the study of superconductivity in atomically thin materials, such as in graphene~\cite{heersche_nature_06,choi_natcom_13,calado_naturenano_15} and transition metal dichalcogenides~\cite{sosenko_arxiv_15,shi_scirep_15}, through the proximity effect from external superconductors. Motivated by this, we set out to determine how superconductivity is manifested in silicene, especially focusing on the external control of unusual phenomena via an electric field.

We demonstrate that proximity-induced superconductivity in silicene allows for a strong quantum ground state control. By creating superconductor-ferromagnet-superconductor (SFS) Josephson junctions in bulk silicene, we find that the exchange field in the F region gives rise to electrically controlled 0-$\pi$ transitions, due to the buckling of the silicene lattice. We also discover that zigzag-oriented SF interfaces, which host notable intrinsic intervalley scattering, result in an exotic $\varphi_0$ state, directly tunable by electric field. Finally, we demonstrate that the existence of Majorana bound states (MBS) at SF junctions on silicene edges is controlled by edge orientation and the strength of electric and in-plane exchange fields.
% ===================== Introduction - END ==================================

\begin{figure}[ht]
  \centering
    \includegraphics[scale=0.47%, bb=0 0 450 450
    ]{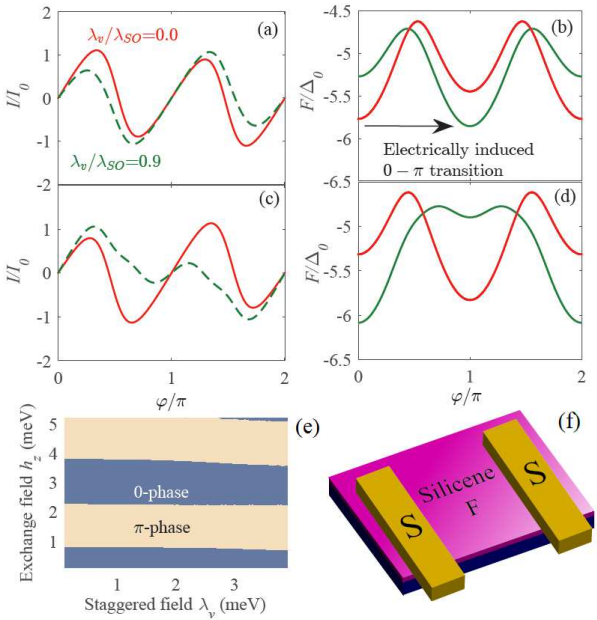}
		  \caption{(Color online.) (a)--(d) Supercurrent and phase-dependent part of the free energy vs superconducting phase difference $\varphi$ for various parameters, illustrating an electric-field-induced 0-$\pi$ transition. We set $I_0=2e\Delta_0/\hbar$. In (a) and (b), $h_z=0.75$ meV and $L_F/\xi_S=0.2$. In (c) and (d), $h_z=5$ meV and $L_F/\xi_S=0.09$, with $L_F$ the junction length. (e) Josephson phase diagram in the $h_z$-$\lambda_v$ plane with $L/\xi_S=0.2$. (f) The proposed experimental setup with superconducting electrodes (yellow) deposited on a silicene sheet (purple) residing on a substrate (blue). Assuming a proximity-induced $\Delta_0 = 0.2$ meV in silicene (a fraction of the supercondcuting order parameter in the host bulk superconductor), the critical supercurrent is of order 50 nA. \label{fig:transition}}
\end{figure}

%========================== 0-pi transition, continuum theory ============================
First, we compute the supercurrent in bulk silicene SFS junctions [see Fig.~\ref{fig:transition}(f)] in the presence of an applied electric field, causing a sublattice staggering. We have done so both in a continuum model and through a numerical lattice calculation, and we proceed to present the continuum results. We use a modified Blonder-Tinkham-Klapwijk formalism~\cite{btk}, adapted to the band structure of a buckled honeycomb lattice. The effective low-energy Hamiltonian near the Dirac points $K$ $(\eta=+1)$ and $K'$ $(\eta=-1)$, incorporating superconducting order $\Delta_0$ and magnetic exchange field $h_z$, reads $\hat{H}_0 = \hat{\rho}_3 m_{\eta\sigma} + v_F(\eta k_x\hat{\rho}_x-k_y\hat{\rho}_y)$ with~\cite{linder_prb_14}
\begin{align*}
\mathcal{H}_{\mathrm{eff}} = \begin{pmatrix}
\hat{H}_0 - (\mu+\sigma h_z)\hat{1} & \sigma\Delta_0\hat{1} \\
\sigma\Delta^{\ast}_{0}\hat{1} & -\hat{H}_0 + (\mu-\sigma h_z)\hat{1}
\end{pmatrix},
\end{align*}
where $\sigma$ is the spin index, $v_F$ the Fermi velocity, $\hat{\rho}_j$ the $j$th Pauli matrix in sublattice space, $\mu$ the chemical potential, $m_{\eta\sigma} = \stpar -\eta\sigma\lambda_\text{SO}$ the mass gap, with the sublattice staggering $\stpar = lE_z$ proportional to an applied electric field, and $\kmpar$ the intrinsic spin-orbit coupling. The basis is $\Psi = [A_{\vk\eta\sigma}^\dag, B_{\vk\eta\sigma}^\dag, A_{-\vk,-\eta,-\sigma}, B_{-\vk,-\eta,-\sigma}]$, with $A (B)$ being the destruction operators on sublattice $A (B)$. To make contact with an experimentally realistic scenario, we assume doped silicene with a chemical potential exceeding the mass gap in order to accommodate proximity-induced superconductivity in the bulk. In the F region, we set the chemical potential to $\mu$ and incorporate an magnetic exchange field $h_z$, induced either via a proximate magnetic insulator or an external magnetic field. The wave functions describing the superconducting regions are: $|\psi_\text{S,left}\rangle$ = $L_e[-\eta\e{\i\beta}, \e{\i\beta},-\eta\e{\i\phi},\e{\i\phi}]\e{-\i q_ex}$ + $L_h[\eta\e{-\i\phi}, \e{-\i\phi}, \eta\e{\i\beta}, \e{\i\beta}]\e{\i q_hx}$ and $|\psi_\text{S,right}\rangle$ = $R_e[\eta\e{\i\beta}, \e{\i\beta},$ $\eta\e{-\i\phi},\e{-\i\phi}]\e{\i q_ex}$ + $R_h[-\eta\e{\i\phi}, \e{\i\phi}, -\eta\e{\i\beta}, \e{\i\beta}]\e{-\i q_hx}$. 
In the central region, we have
\begin{align}
\left\vert\psi\right\rangle &= 
\begin{bmatrix}
a(\eta k_x^+ + \i k_y)\e{\i k_x^+x} + b(-\eta k_x^+ + \i k_y)\e{-\i k_x^+x} \\
aQ_+\e{\i k_x^+x} + bQ_+\e{-\i k_x^+x} \\
c(-\eta k_x^- -\i k_y)\e{\i k_x^-x} + d(\eta k_x^- -\i k_y)\e{-\i k_x^-x}\\
cQ_-\e{\i k_x^-x} + d Q_-\e{-\i k_x^-x}
\end{bmatrix} 
\end{align}
with $k_x^\pm = \sqrt{(\mu \pm \sigma h_z)^2 - m_{\eta\sigma}^2 - k_y^2}$ and $Q_\pm = \sigma h_z \pm (\mu + m_{\eta\sigma})$. Above, $\beta=\arccos(\varepsilon/\Delta_{0})$ with $\varepsilon$ the energy, $\pm\phi$ the superconducting phase on the left/right side, while $\{L_{e,h},R_{e,h},a,b,c,d\}$ are the scattering coefficients for each quasiparticle excitation. When the superconducting regions are strongly doped, the Fermi vector mismatch leads to transport predominantly occurring via normal incidence of the quasiparticles, and we focus on this regime.
By connecting the wave functions $|\psi_\text{S}\rangle$ and $|\psi\rangle$ at their respective SF interfaces, we find an analytical condition for the existence of non-trivial solutions of the scattering coefficients. The energies $\varepsilon$ that solve this equation are the Andreev bound states, which carry the supercurrent through the junction. We find a set of bound states of the form $\varepsilon_{\eta\sigma}^\pm = \Delta_0\sqrt{[C_1(\varphi) \pm \sqrt{C_2(\varphi)}]/C_3} \in [0,\Delta_0]$, where the coefficients $C_1$ and $C_2$ depend on the superconducting phase difference $\varphi\equiv 2\phi$, while $C_3$ does not~\cite{SM}. With the bound state spectrum in hand, the supercurrent and phase-dependent part of the free energy are $I_J = $$ -\frac{2e}{\hbar} \sum_{\pm,\eta,\sigma} \text{tanh}(\varepsilon^\pm_{\eta\sigma}/2k_BT)$$(d\varepsilon^\pm_{\eta\sigma}/d\varphi),$ $F = $$-1/k_B T\sum_{\pm,\eta,\sigma}$$[\ln(1\!+\!\e{-\varepsilon_{\eta\sigma}^\pm/k_B T}) + \ln(1+\e{\varepsilon_{\eta\sigma}^\pm/k_BT})]$.
%========================== 0-pi transition, continuum theory - END ==============================

%====================== 0-pi transition, continuum Results - BEGIN ===============================
We first demonstrate that the present system offers the possibility to tune the superconducting quantum ground state and induce 0-$\pi$ transitions electrically by changing the applied external field $E_z$. With a conventional BCS superconductor comprising the leads, an induced gap of $|\Delta_0| \sim 0.2$ meV in silicene is reasonable, which is  notably less than a bulk superconducting gap of order $1$ meV. The spin-orbit coupling $\lambda_\text{SO} = 3.9$ meV is fixed and we set $T/T_c=0.2$. In Figs.~\ref{fig:transition}(a)--\ref{fig:transition}(d) we show that for two representative exchange fields $h_z$ and junction lengths, a 0-$\pi$ transition is obtained by simply turning on an external electric field $E_z = 0.9\lambda_\text{SO}$. Such electric control over the superconducting quantum state is very different from conventional metallic SFS junctions, where, e.g., either the temperature of a given sample needs to be changed incrementally~\cite{ryazanov_prl_01} or several samples with different lengths have to be manufactured~\cite{oboznov_prl_06} in order to observe this feature. In the present case, the 0-$\pi$ transition is simply controllable \textit{in situ} via an electric field. 

The physical origin of the quantum ground state control via the electric field $E_z$ is that the total phase shift picked up by the quasiparticles that constitute the Andreev bound state in silicene is tuned by $E_z$ in a spin-dependent manner. This can be seen from the form of the wave vectors $k_x^\pm$. While spin-$\sigma$ electrons from valley $\eta$ are degenerate with $-\sigma$ electrons from the opposite valley $-\eta$ in the absence of an exchange field, this degeneracy is lifted when $h_z\neq 0$. In this case, any change in the electric field will affect one spin carrier in a way that cannot be compensated for by switching both the spin and valley quantum numbers. As a result, the electric field tunes the spin-dependent phases picked up by the Andreev bound state quasiparticles traversing the junction, thus triggering a 0-$\pi$ transition. Changing $h_z$ is more effective in triggering 0-$\pi$ transitions than altering $E_z$. The reason for this is that the phase accumulated by the Andreev bound state has one contribution of order $O(h_z/\mu)$ and a different contribution of order $O(h_z/\mu \times m^2/\mu^2)$, with $m$ the mass gap tunable by $E_z$. Since both $h_z/\mu$ and $m/\mu$ are both small numbers, it takes a larger increase in $E_z$ to result in a net $\pi$ addition to the phase, which is consistent with the phase diagram in Fig.~\ref{fig:transition}(e).
%====================== 0-pi transition, continuum Results- END =================================

%===================== Lattice theory - BEGIN ========================
To confirm these 0-$\pi$ transition findings, we have performed complementary lattice calculations using the Bogoliubov de Gennes (BdG) framework. The advantages of a real-space lattice approach over continuum calculations are in automatically incorporating the correct atomistic boundary conditions, taking interface orientations, intervalley scattering, and edge effects into account, as well as allowing for self-consistent calculations. Due to computational constraints, lattice calculations are, however, limited to using relatively small superconducting coherence lengths $\xi_S$. Nevertheless, by scaling spatial dimensions by this quantity, good quantitative predictions can still be obtained~\cite{SM,grapheneJJ,grapheneJJexp}.
As appropriate for silicene, we use a tight-binding model on the honeycomb lattice for the normal state Hamiltonian~\cite{kanemele2005graphene,Liu11},
\begin{align}
\label{eq:matH0}
\mathcal{H}_{0}  & = t \sum_{\langle i, j\rangle\alpha} c^{\dagger}_{i\sigma} \, c_{j\sigma} + \frac{i \kmpar}{3\sqrt{3}} \sum_{\langle\langle i, j \rangle\rangle\alpha\beta} \nu_{ij} (\sigma_{z})_{\alpha\beta} \, c^{\dagger}_{i\alpha} \, c_{j\beta} \nonumber \\
&  -  \sum_{i\alpha} \tilde{\mu}_{i} \, c^{\dagger}_{i\alpha} \, c_{i\alpha},
\end{align}
where $c_{i\sigma}^\dagger$ is the creation operator on site $i$, $t$ the nearest-neighbor hopping parameter, $\nu_{ij} = \pm 1$ depending on a counterclockwise or clockwise turn when hopping from $j$ to next-nearest-neighbor site $i$, and $\tilde{\mu}_{i} = \mu + \stpar \zeta_{i}$, with $\zeta_{i} = \pm 1$ for the two different sublattices.
The applied exchange field in the F region is described by $\mathcal{H}_{F} = -\sum_{i\alpha\beta} (\bm{h}_{i} \cdot \bm{\sigma})_{\alpha\beta} \, c^{\dagger}_{i\alpha} \, c_{i\beta}$, allowing for both out-of-plane $h_{z}$ and in-plane $h_{\|}$ components. Superconductivity is induced by proximity to conventional $s$-wave superconducting contacts and is modeled by the term $\mathcal{H}_{\Delta} = \sum_{i} \Delta_{i} c^{\dagger}_{i\su} \, c^{\dagger}_{i\sd} + \mathrm{H.c.}$, where $\Delta_{i}$
is calculated self-consistently using $\Delta_i = -U_i \langle c_{i\sd}c_{i\su}\rangle$~\cite{selfcons,SFS_QSHI}. Here the effective on-site attractive interaction $U_i$ is proximity induced only in the S regions~\cite{lattice}. Note that a self-consistent treatment of $\Delta_i$ is necessary for capturing the full influence of the silicene electronic structure on the superconducting state. To establish the superconducting phase dependence, we fix the phase of $\Delta_i$ to $0$ and $\varphi$, respectively, in the two S regions. We find the ground state by calculating the free energy, or the grand thermodynamic potential, difference between the superconducting $F$ and the normal state $F_0$ for different $\varphi$~\cite{SM}. 

%===================== Lattice theory - END ========================

\begin{figure}[hb]
\includegraphics[scale=1.0]{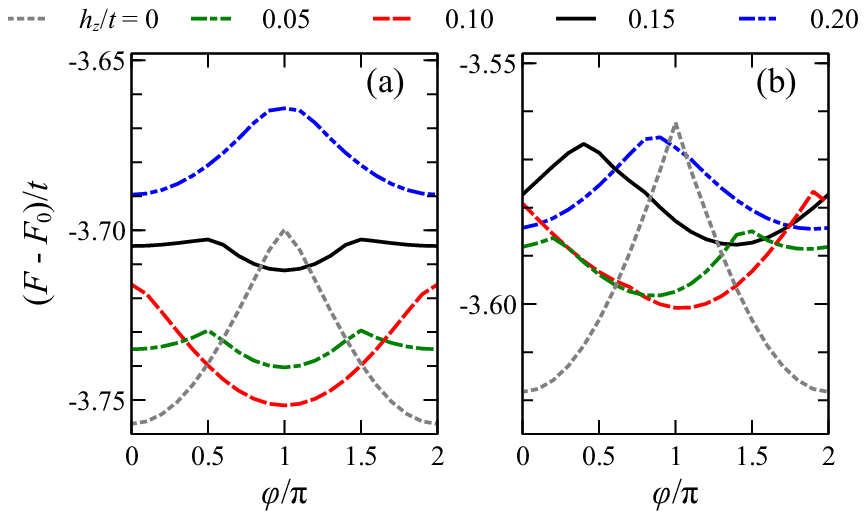}
\caption{\label{fig:ZZSelfCons} Free-energy difference $F-F_0$ as a function of superconducting phase difference $\varphi$ for a SFS junction with ZZ interfaces with staggering (a) $\stpar/\kmpar = 0.0$ and (b) $\stpar/\kmpar = 1.0$ for varying exchange field $0.0 \le h_{z}/t \le 0.2$ (F region only). Here, $\kmpar/t = 0.2$, $\mu/t = 0.5$, $U/t = 2.43$ (S regions only, giving $\xi_S/L_S \approx 0.25$). Each region is $20$ (9) atoms wide in the AC (ZZ) direction.}
\end{figure}

%===================== varphi_0 results - BEGIN ========================
Using the lattice BdG approach we have confirmed the continuum results, finding a sequence of 0-$\pi$ transitions driven by both electric and exchange fields.
Surprisingly, we also find that for zigzag (ZZ) interfaces for the superconducting contacts, a very notable $\varphi_0$ ground state appears with increasing electric fields, as clearly displayed in Fig.~\ref{fig:ZZSelfCons}.
When the electric field is absent, $\stpar = 0$, the free-energy phase dependence is symmetric with respect to $\varphi = 0$ (or $\varphi = \pi$). Then, with increasing $h_{z}$ the system undergoes a series of $0$-$\pi$ transitions [see Fig.~\ref{fig:ZZSelfCons}(a)], as expected from the continuum results. However, with a finite $\stpar$, the phase dependence also develops a clear asymmetry, with a notable component that is odd in $\varphi$ [see Fig.~\ref{fig:ZZSelfCons}(b)]. Thus, the lowest free energy is achieved for a phase that is not $0$ or $\pi$, but an arbitrary value $\varphi_{0}$. Thus, both time-reversal and sublattice symmetry breaking, by $h_z$ and $\stpar$, respectively, are necessary for the appearance of an $\varphi_0$ state~\cite{zazunov_prl_09,phi_experiment}. 

We find that the most important factor contributing to the asymmetry of the phase dependence is the choice of interface. For a ZZ interface the asymmetry is very prominent, whereas it is always very minor for armchair (AC) interfaces. The main difference between the two interfaces is in the significant intervalley scattering present at ZZ interfaces, for which the Dirac cones from the two valleys project on top of each other. 
No intervalley scattering is included in the continuum model, as this rendered the problem analytically untractable, which is why the $\varphi_0$ state does not appear there. In contrast, the numerical lattice calculations show that the coupling of the valleys in the ZZ case results in a very notable $\varphi_0$ phase shift of the current-phase relation.
This mechanism is different from previous proposals on how to realize $\varphi_0$-junctions, which include quantum dots~\cite{zazunov_prl_09}, as recently experimentally observed~\cite{phi_experiment}, the combination of Rashba spin-orbit coupling and suitably oriented magnetic fields~\cite{buzdin_prl_08,liu_prb_10,yokoyama_prb_14}, chiral spin configurations \cite{eschrig_prl_09,chan_prb_10,kulagina_prb_14}, and topological insulator surface states~\cite{tanaka_prl_09,SFS_QSHI}. Specifically, the $\varphi_0$ state in bulk silicene is not only sensitive to interface orientation but is also easily controlled by applying an electric field.

Beyond intervalley scattering, we find that some asymmetry is also induced by lattice effects on the superconducting order parameter. Particularly, lattice staggering in the S regions, as well as the inverse proximity effect, cause asymmetry, effects that are also ignored in the continuum calculations. One measure of the asymmetry is the magnitude of the odd-$\varphi$ part of the free energy. Removing $\stpar$ in the S regions and fixing $\vert \Delta \vert$ to a constant value, with only a constant phase difference $\varphi$ between the two contacts, eliminates both the staggering effects on the order parameter and the inverse proximity effect. This decreases the odd part of the free energy for both the ZZ and AC interfaces. For the AC interface the odd part is reduced by an order of magnitude and reaches the numerical accuracy limit of our calculations. For the ZZ interface the reduction is less dramatic and asymmetry is still clearly present~\cite{SM}.

%================  Majorana - BEGIN  ====================================
So far we have concentrated on the doped metallic regime of bulk silicene. Pristine silicene is also very interesting as it is a quantum spin Hall insulator (QSHI), with a bulk band gap but topologically protected fully spin-polarized metallic edge states~\cite{kanemele2005graphene,kanemele2005QSHI,liu_prl_11}. Here, we explore if the silicene QSHI edge states harbor zero-energy Majorana bound states (MBS) at SF interfaces and especially their dependence on electric field. For this purpose, we set $\mu = 0$ and the value of $U$
%strength of the Hubbard-$U$ interaction
such that superconductivity is only induced in the metallic silicene-vacuum edge states. 
The proximity-induced superconducting edge states form a quasi-one-dimensional (1D) gapped topological superconductor. This topological state has been shown to host single MBS when interfaced with a F region, both in the continuum limit~\cite{FuKane2009} and for a prototype honeycomb QSHI with ZZ edges~\cite{SFS_QSHI}. The requirement for MBS is that the F region exchange field is not parallel to the spin-quantization axis of the normal state, otherwise spin-rotation symmetry prevents single MBS~\cite{SM}. Here, we find that for AC silicene edge states, exchange fields in any in-plane direction, necessary for silicene MBS, fail to gap the metallic edge states.
Thus, an SF junction along the AC silicene edge cannot host MBS, since they will necessarily hybridize with the metallic states in the F region~\cite{SM}.

%================ Soft Majorana - BEGIN  ====================================
Therefore, we only investigate the evolution of MBS at SF junctions along ZZ edges with increasing staggering $\stpar$. The bulk gap closes in silicene when $\stpar = \kmpar$ and larger staggering gives a trivial insulator~\cite{kanemele2005QSHI,ezawa_njp_12}. Thus, we expect that any MBS will be lost at this critical staggering. However, we find that the existence of MBS is also heavily dependent on the exchange field in the F region. 
\begin{figure}[ht]
\includegraphics[scale=1.0]{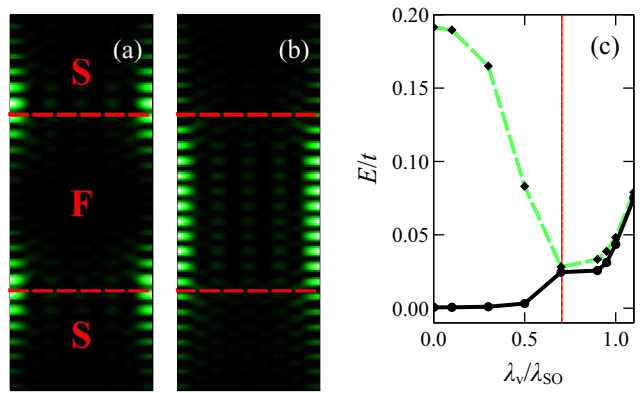}
\caption{\label{fig:SoftMajor} Probability density of the in-gap BdG eigenvectors for (a) $\stpar/\kmpar = 0.0$ and (b) $\stpar/\kmpar = 0.7$, with white/green representing high densities. (c) Energy of the only in-gap state (solid black) and the second lowest-energy state, corresponding to bulk states (dashed green) vs staggering, with the red vertical line marking the closing of the edge state gap in the F region. The SFS junction has ZZ vacuum edges with proximity-induced superconductivity only in the edge states. Here, $\kmpar/t = 0.5$, $\mu = 0$, $U/t = 2.0$ (S regions only) with $\varphi  = 0$, and $h_{\|}/t = 0.3$ (F region only). Each region is $20$ ($12$) atoms wide in the AC (ZZ) direction.}
\end{figure}
Figures~\ref{fig:SoftMajor}(a) and \ref{fig:SoftMajor}(b) illustrate the case of a weak exchange field in a F region between two S regions, such that $h_{\|} < \kmpar$. As staggering $\stpar$ is increased, the two MBS sitting on the same edge start hybridizing along the edge in the F region, with no notable spread into the bulk. Interestingly, this happens for small enough staggering that silicene is still well within the QSHI phase. The reason for this is that, while $h_{\|}$ opens a mass gap in the edge states on the ZZ edge, $\stpar$ acts to decrease it, eventually closing it. When this happens, any in-gap states hybridize with the continuum states in the F region and the zero-energy MBS are destroyed, as clearly depicted in Fig.~\ref{fig:SoftMajor}(c). Thus, for weakly ferromagnetic junctions, the existence of MBS is highly tunable even with small electric fields.

%===================== Hard Majorana - BEGIN ================================
\begin{figure}[ht]
\includegraphics[scale=1.0]{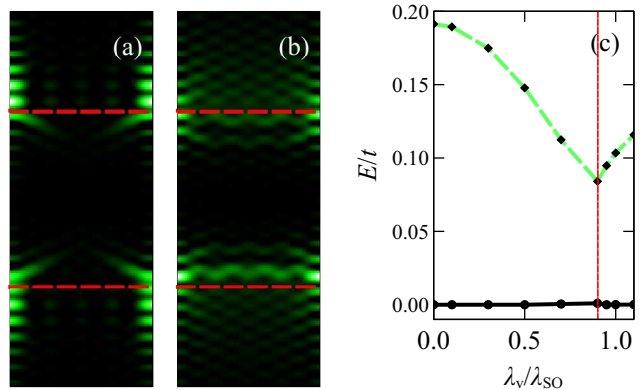}
\caption{\label{fig:HardMajor} Same as Fig.~\ref{fig:SoftMajor} but with (a) $\stpar/\kmpar = 0.0$ and (b) $\stpar/\kmpar = 1.1$, and $h_{\|}/t = 0.7$ (F region only). The bulk energy state in (c) does not reach zero at the topological phase transition (red vertical line) due to a finite sized sample.}
\end{figure}
In contrast, Fig.~\ref{fig:HardMajor} presents the case for strong exchange fields, such that $h_{\|} > \kmpar$. In this case, the edge state energy gap in the F region does not close before staggering destroys the QSHI state. This leads to the MBS being preserved until this critical staggering, at which point the energy gap in the bulk closes, due to the topological phase transition, and the MBS start hybridizing along the SF interface, across to the other edge of the finite ribbon, creating a regular electronic quasiparticle excitation.
Remarkably, even though superconductivity is completely suppressed for $\stpar> \kmpar$, since the edge states have disappeared and the bulk is insulating, this electronic state remain at zero energy, as seen in Fig.~\ref{fig:HardMajor}(c). 
Two MBS in the superconducting phase thus evolve into one zero-energy electron quasiparticle state in the nonsuperconducting state, localized along the SF interface across the ribbon.
%====================== Hard Majorana - END ---===============================

In conclusion, we have discovered both $0$-$\pi$ and $\varphi_0$ states in doped bulk silicene SFS junctions, which are directly controllable by an electric field. We have also found that the existence of MBS at silicene edges is strongly dependent on electric field. It is the unique combination of a honeycomb lattice, spin-orbit coupling, and buckling directly tunable by electric field that results in this exceptional amount of ground state control in silicene Josephson junctions. By analogy, Josephson junctions in the recently discovered materials germanene and stanene~\cite{Liu11,Ge_exp,Sn_exp} should display the same properties.

\begin{acknowledgments}
We thank T.~Yokoyama and A.~Bouhon for valuable discussions. D.K. and A.B.-S. acknowledge funding from Carl Trygger's Foundation, the G\"oran Gustafsson Foundation, the Swedish Research Council (Vetenskapsr\aa det), the Swedish Foundation for Strategic Research (SSF), and the Wallenberg Academy Fellows program through the Knut and Allice Wallenberg Foundation. J.L. acknowledges funding via the Outstanding Academic Fellows program at NTNU, the COST Action MP-1201 and the Research Council of Norway Grants No. 205591, No. 216700, and No. 240806.
\end{acknowledgments}

\clearpage
\onecolumngrid{
\subsection{Detailed expression for the Andreev bound state energies}
We find the following analytical expression for the Andreev bound state energies:
\begin{align}
\varepsilon_{\eta\sigma}^\pm = \Delta_0\sqrt{[C_1(\varphi) \pm \sqrt{C_2(\varphi)}]/C_3} \in [0,\Delta_0],
\end{align}
with
\begin{align}
C_1 = 2(a^2+b^2)-aB,\; C_2 = b^2(4a^2+4b^2-B^2),\; C_3 = 4(a^2+b^2).
\end{align}
Above, we have defined the following quantities:
\begin{align}
a &= 4(q_+^2 + k_+^2)(q_-^2+k_-^2)\sin(k_+L)\sin(k_-L) - 16q_+q_-k_+k_-\cos(k_+L)\cos(k_-L),\notag\\
b &= 8(q_+^2+k_+^2)q_-k_-\sin(k_+L)\cos(k_-L) + 8(q_-^2+k_-^2)q_+k_+\cos(k_+ L)\sin(k_-L),\notag\\
c &= 64q_+q_-k_+k_-\cos^2(\varphi/2)-8\sin(k_+L)\sin(k_-L)(q_+^2-k_+^2)(q_-^2-k_-^2)-32q_+q_-k_+k_-.
\end{align}
Here, $L$ is the length of the junction, while
\begin{align}
k_\pm = \sqrt{(\mu\pm\sigma h_z)^2 - (\eta\sigma\lambda_\text{SO}-\stpar)^2},\; q_\pm = \pm \mu + \sigma h_z \pm (\eta\sigma\lambda_\text{SO}-\stpar).
\end{align}

\subsection{Lattice calculation details}
\subsubsection{Phase dependence of the free energy}
For the lattice calculations of the phase dependence of the free energy, we consider a silicene SFS heterojunction on a finite-sized sample, with a relative phase of the superconducting order parameter between the two S regions. We arrange the SFS junction such that its SF interfaces are along either the ZZ or AC directions of the honeycomb lattice.

Due to computational limitations, the overall number of atoms can be no larger than approximately $4 \times 141$. We choose the sample size to be $20 \times (9 + 9 + 9)$ atomic lines for ZZ-interfaced SFS junctions (i.e.~same nine-atomic-line wide regions for each S and F region), and $(20 + 20 + 20) \times 9$ atomic lines for AC-interfaced junctions. Because the inter-atomic distances are different for ZZ and AC directions, the actual geometric size of each region is approximately $14.0 \times (15.2 + 15.6 + 15.2)$ units of silicene nearest-neighbor distance in the ZZ junctions, and $(14.5 + 15.0 + 14.5) \times 14.7$ units in the AC junctions. These are the closest dimensions to square regions that can be realized on the honeycomb lattice, given the constraint on the total number of atoms in the sample.

In order for the superconducting state to reach bulk conductions in the middle of each S region, we need the superconducting coherence length ($\xi = \hbar \, v_{F}/\vert \Delta_0 \vert$, where $\hbar \, v_{F} = 3 t \, a/2$) to be around one quarter of the length of the S region. This results in the superconducting order parameter being relatively large at $\vert\Delta\vert/t = 0.41$, which is essentially the only weakness of a lattice BdG approach. But, by scaling all spatial dimensions by the superconducting coherence length $\xi_S$, as we do here, it is still possible to make qualitative as well as quantitative predictions. One recent instance where it was clearly shown that this lattice BdG method gives very good experimental predictions is graphene Josephson junctions, where calculated current-phase relations~\cite{grapheneJJ} match the experimentally measured data extremely well~\cite{grapheneJJexp}. These graphene calculations used a very similar model, method, and system sizes, as the current study, which validates our study.

For studying the properties of bulk silicene SFS junctions, with their the $0$-$\pi$ transitions and $\varphi_0$ states, we choose the chemical potential to be in the middle of the energy band, between the Dirac point at $\mu = 0$ and the van Hove singularity at $\mu = t$, by fixing $\mu = 0.5 \, t$ in the whole sample. In order to be consistent with the size of the superconducting order in the BdG approach, we also need to scale the spin-orbit coupling, such that both terms reflect the same relative size of their induced energy gaps, as expected experimentally. For this purpose, we set $\kmpar = 0.2 t$, to reflect the same relative sizes. 
We have also here ignored the much smaller Rashba spin-orbit coupling~\cite{ezawa_njp_12,Liu11}, as we do not expect it to notably influence our results.
With these values, we find that $U = 2.43 \, t$ is the necessary value in order to produce the required superconducting bulk system. For the staggering $\stpar$, driven by an applied external electric field, we study three different values $\stpar/\kmpar = 0.$ (no staggering), $0.5$, and $1.0$ (critical staggering resulting in the topological phase transition in the QSHI phase), although we only report results for the first and third values. Similarly, we vary the exchange field $h_{z}$ in the F region in the interval $0.0 \le h_{z}/t \le 0.2$, while keeping its direction is perpendicular to the plane of the silicene sheet for all phase-dependent free-energy studies. Here $z$ is the good spin-quantization direction for the spin-orbit coupling.

The self-consistent calculation proceeds by choosing as the initial guess for the superconducting order parameter the corresponding bulk value in an infinite system, with the phase difference between the two S regions fixed to $\varphi$. Then, the superconducting order parameter is reiteratively calculated, until the maximal difference in the magnitude of the superconducting order parameter at any lattice point of the sample between two successive iterations is smaller than a predefined tolerance set to $0.5 \times 10^{-4} \, t$. The free energy (or rather the grand thermodynamic potential) is then calculated for the converged configuration of the superconducting order parameter and compared to the free energy of the normal state.

\subsubsection{Majorana bound states}
For the MBS study, we are required to be in the QSHI phase with the chemical potential in the bulk energy gap. We therefore set $\mu = 0$ throughout the sample, which corresponds to pristine silicene. Then, in order to emphasize the in-gap states, we exaggerate the bulk gap, due to the spin-orbit coupling, by setting $\stpar = 0.5 \, t$. The Hubbard-$U$ value in the S regions is set to $U = 2.0 \, t$, which is sufficient to induce a stable superconducting gap in the edge states, but does not render the bulk superconducting, even at the topological phase transition. For these calculations, the sample size is set to $20 \times (12 + 12 + 12)$ for the ZZ (Main Text Figs.~3, 4) and $(36 + 36 + 36) \times 8$ atomic lines for the AC interfaces (Supplementary Fig.~\ref{fig:NoMajorArmChair}).

Due to the particle-hole symmetry of the BdG Hamiltonian; if $\left( u^{(n)}_{i\sigma} \vert v^{(n)}_{i\sigma} \right)^{\top}$ is an eigenvector of the BdG Hamiltonian corresponding to an energy $E_{n} \ge 0$, then $\left( (v^{(n)}_{i\sigma})^{\ast} \vert (u^{(n)}_{i\sigma})^{\ast} \right)^{\top}$ is also an eigenvector corresponding to energy $E_{n} \le 0$. The eigenvector with $E_{n} > 0$ defines a quasiparticle excitation $\Gamma^{\dagger}_{n} = \sum_{i\sigma} u^{(n)}_{i\sigma} \, c^{\dagger}_{i\sigma} + v^{(n)}_{i\sigma} \, c_{i\sigma}$, while the particle-hole symmetric one defines the corresponding annihilation operator $\Gamma_{n}$. Then, when the energy $E_n = 0$, the creation and annihilation operator define a degenerate excitation, and the linear combinations $\gamma_{2n-1} = e^{i \, \alpha_{n}} \, \Gamma_{n} + e^{-i \, \alpha_{n}} \, \Gamma^{\dagger}_{n}$, and $\gamma_{2n} = i (\, e^{i \, \alpha_{n}} \, \Gamma_{n} - e^{-i \, \alpha_{n}} \, \Gamma^{\dagger}_{n})$, where $\alpha_{n}$ are arbitrary phases that may be absorbed in a re-definition of $\Gamma_{n}$, satisfy the Majorana condition:
\[
\gamma_{a} = \gamma^{\dagger}_{a}, \ \left\lbrace \gamma_{a}, \gamma_{b} \right\rbrace = 2 \, \delta_{ab}, \ (a, b = 1, 2, \ldots, 2n).
\]
Thus, zero-energy states in a superconductor generally fulfill the Majorana condition. However, unless it is possible to spatially separate the two zero-energy solutions into two distinct single solutions, these two states just join and form a regular Bogoliubov (electron-like as compared to Majorana-like) quasiparticle excitation. 
Technically, a spatial separation is enabled by breaking the spin-rotation symmetry of the BdG Hamiltonian, which forces each lattice site to require a four-component Nambu vector. This is exactly the artificial doubling of the electronic degrees of freedom required for single and spatially-well-separated MBS solutions.
For a silicene SFS junction this requires an exchange field in the F region with an in-plane component $h_{\|}$, since the normal state has a good spin-quantization axis along the out-of-plane $z$-axis. 
Such a SFS junction along a single ZZ silicene edge results in two MBS; one localized at each SF interface. Since we model silicene nanoribbons, we have two ZZ silicene edges, and, thus, a total of four MBS. These four zero-energy states are localized at each of the four SF interfaces along the silicene edges, and since all four combined only comprise two Bogoliubov quasiparticles, they are the condensed matter physics incarnation of four Majorana fermions. We here choose to plot $\sum_{n, \sigma} \vert u^{(n)}_{i\sigma}\vert^{2} + \vert v^{(n)}_{i\sigma}\vert^{2}$ for all in-gap states , since the MBS are the only states inside the gap, and this directly displays the location of each MBS. It has the added advantage that, when the energy of this lowest in-gap state drifts away from zero, we can still follow the spatial distribution, although, it does not anymore correspond to a MBS excitation, or even a localized excitation of the system.
Notably, once two MBS hybridize either along the edge through the F region, or along the SF interface across the ribbon, they are reduced to a single electronic-like quasiparticle, regardless if the energy stays at zero or not.

\subsection{Additional numerical lattice BdG results}
\subsubsection{Self-consistency effects on the $\varphi_0$ state}
Main Text Figure.~2 and Supplementary Fig.~\ref{fig:ACSelfCons} display the results of fully self-consistent calculations for SFS junctions with ZZ and AC interfaces, respectively, with staggering $\stpar$ present in all regions. As mentioned in the main text, the phase asymmetry of the free energy is heavily suppressed in the case of AC interfaces.
\begin{figure}[h]
\includegraphics[width=0.5\linewidth]{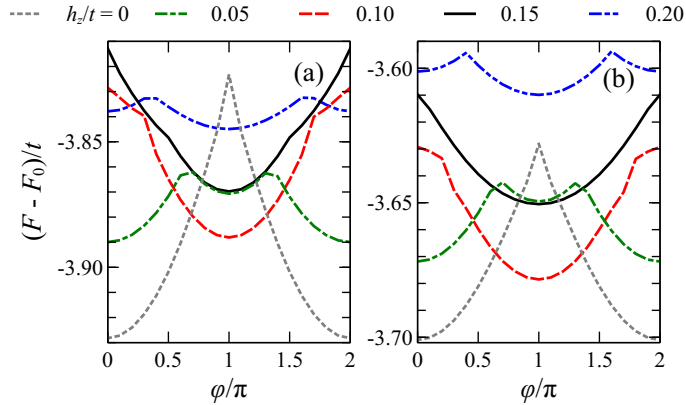}
\caption{\label{fig:ACSelfCons}Free-energy difference $F-F_0$ as a function of superconducting phase difference $\varphi$ for an SFS junction with AC interfaces with staggering (a) $\stpar/\kmpar = 0.0$ and (b) $\stpar/\kmpar = 1.0$ for varying exchange field $0.0 \le h_{z}/t \le 0.2$ (F region only). Here, $\kmpar/t = 0.2$, $\mu/t = 0.5$, $U/t = 2.43$ (S regions only, giving $\xi_S/L_S \approx 0.25$). Each region is 20 (9) atoms wide in the AC (ZZ) direction. Same figure as Main Text Fig.~2 but for AC interfaces.}
\end{figure}
Although not visible for the AC interface in Supplementary Fig.~\ref{fig:ACSelfCons}, there is, however, still a small part of the free energy that is odd in $\varphi$ even for AC interfaces, but the maximum value is two orders of magnitude smaller than the corresponding case for an ZZ interface. The exact values are summarized in Supplementary Table~\ref{table:oddparts}.
\begin{table}[h]
%\begin{ruledtabular}
\begin{tabular}{||c|cc|cc||}
\hline \hline
 & \multicolumn{2}{c|}{self-consistent} & \multicolumn{2}{c||}{non self-consistent} \\
 & $\stpar/\kmpar = 0.0$ & $1.0$ & $0.0$ & $1.0$ \\ 
 \hline
 ZZ & $0$ & $1.0\times 10^{-2}$ & $0$ & $8.0\times 10^{-3}$ \\
 \hline
 AC & $2.8\times 10^{-4}$ & $3.8\times 10^{-4}$ & $4.2\times 10^{-5}$ & $3.4\times 10^{-5}$ \\
\hline \hline
\end{tabular}
%\end{ruledtabular}
\caption{\label{table:oddparts} Maximal values for the odd-$\varphi$ part of the free energy, $(F(\varphi) - F(-\varphi))/(2t)$, for the cases presented in Main Text Fig.~2, Supplementary Figs.~ \ref{fig:ACSelfCons}--\ref{fig:ACnoSelfCons}.}\end{table}

In order to control for the role of $\stpar$ in the superconducting regions and the inverse proximity effect captured by a fully self-consistent calculation, both of which are absent in the continuum model, we have also performed additional restricted calculations. Here we use $\stpar = 0$ in the S regions and set $\vert \Delta \vert$ to a fixed magnitude, equal to the self-consistent value in the bulk, but still with the phase difference fixed to $\varphi$ between the two S regions. 
Supplementary Figure~\ref{fig:ZZnoSelfCons} displays the result for a non self-consistent calculation with no staggering in the S regions for a ZZ-interfaced SFS junction. Panel (b) is still clearly asymmetric for finite electric fields, with the maximum values for the odd-$\varphi$ somewhat reduced, but non-vanishing, see also Supplementary Table~\ref{table:oddparts}.
\begin{figure}
\includegraphics[scale=1.0]{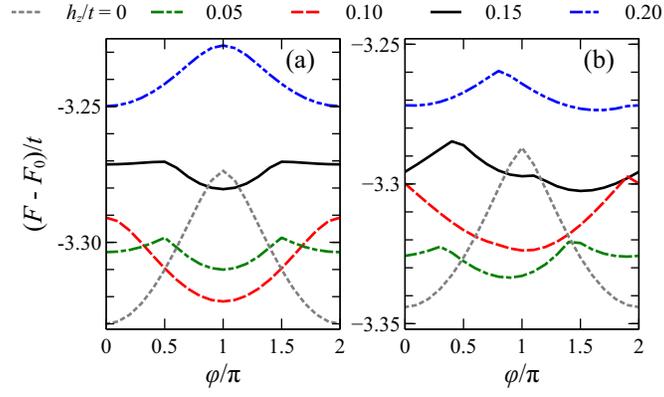}
\caption{\label{fig:ZZnoSelfCons} Same figure as in Main text Fig.~2 but with no lattice staggering in the S regions and fixing $\vert \Delta \vert / t = 0.41$ in the S regions, without performing a self-consistent calculation.}
\end{figure}

Finally, Supplementary Fig.~\ref{fig:ACnoSelfCons} displays the result for a non self-consistent calculation with no staggering in the S regions for a SFS junction with AC interfaces. Here the odd-$\varphi$ parts are reduced by an order of magnitude [cf. Supplementary Table~\ref{table:oddparts}], resulting in values that are clearly below the numerical tolerance of the self-consistency calculations. 
The AC case with a non self-consistent superconducting order parameter is the setup most equivalent with the continuum model, and it also displays the same results as the continuum model. 
\begin{figure}[h]
\includegraphics[scale=1.0]{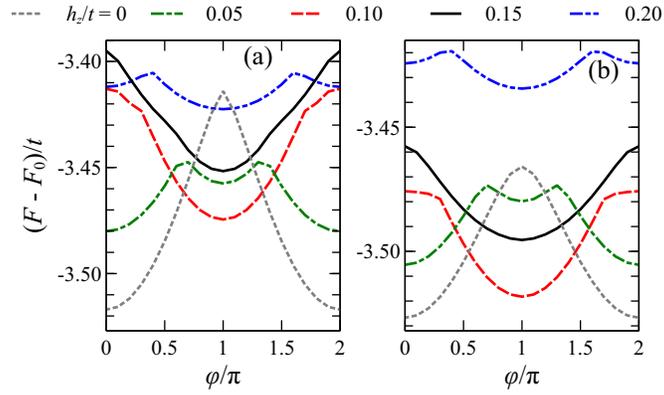}
\caption{\label{fig:ACnoSelfCons} Same figure as in Supplementary Fig.~\ref{fig:ZZnoSelfCons} but for AC interfaces.}
\end{figure}

\subsubsection{Lack of MBS for AC edges}
In the case of a finite silicene sample with zero or low doping, such that the chemical potential is within the bulk energy gap, there are topologically protected metallic edge states. A finite in-plane exchange field $h_{\|}$ has previously been shown to open an energy gap in these edge states in a generic continuum model~\cite{FuKane2009} as well as along ZZ edges~\cite{SFS_QSHI}. However, we find that for AC edges a $h_{\|}$ field does {\it not} gap the metallic edge states, independent on the in-plane direction. Therefore, the lowest energy state for a SFS junction along the AC edge is spread through the whole edge of the F region, not allowing for the existence of zero-energy MBS localized at the SF interfaces, as clearly depicted in Supplementary Fig.~\ref{fig:NoMajorArmChair}.  
\begin{figure}[ht]
\includegraphics[scale=1.0]{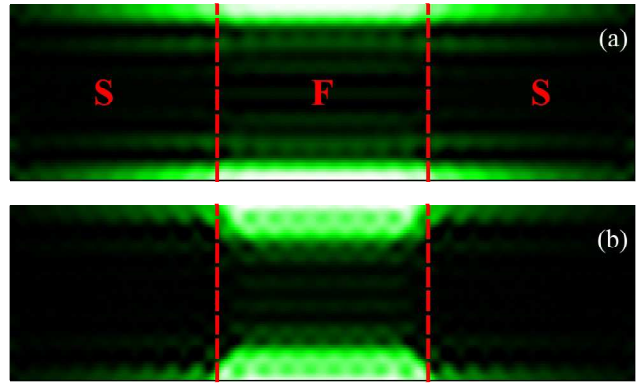}
\caption{\label{fig:NoMajorArmChair}
Probability density of the in-gap BdG eigenvectors for (a) $h_{\|} = 0.3$ and (b) $h_{\|}/t = 0.7$ (F region only), with white/green representing high densities.
The SFS junction has AC vacuum edges with superconductivity only induced on the edge states. Here, $\kmpar/t = 0.5$, $\mu = 0$, $\stpar/t = 0$, and $U/t = 2.0$ (S regions only) with $\varphi  = 0$. Each region is $36$ (8) atoms wide in the AC (ZZ) direction.}
\end{figure}}
\clearpage
\end{document}